\documentclass[apjl]{emulateapj}

\include{vc}

\usepackage[english]{babel}
\usepackage{amsmath}
\usepackage{graphicx}
\usepackage[colorinlistoftodos]{todonotes}

\shortauthors{Barclay et al. 2013}
\shorttitle{Kepler-91\lowercase{b} is a giant planet orbiting a giant star}

\begin{document}
\title{Radial velocity observations and light curve noise modeling confirm that Kepler-91\lowercase{b} is a giant planet orbiting a giant star\footnotemark[1]}
\footnotetext[1]{Based partly on observations obtained with the Hobby-Eberly Telescope, which is a joint project of the University of Texas at Austin, the Pennsylvania State University, Stanford University, Ludwig-Maximilians-Universit\"{a}t M\"{u}nchen, and Georg-August-Universit\"{a}t G\"{o}ttingen.}

\author{Thomas Barclay$^{2,3}$, Michael Endl$^{4}$, Daniel Huber$^{2,5}$,  Daniel Foreman-Mackey$^{6}$, William D. Cochran$^{4}$, Phillip J. MacQueen$^{4}$, Jason F. Rowe$^{2,5}$ and Elisa V. Quintana$^{2,7}$}

\slugcomment{Accepted for publication in ApJ}

\altaffiltext{2}{NASA Ames Research Center, M/S 244-30, Moffett Field, CA 94035, USA}
\altaffiltext{3}{Bay Area Environmental Research Institute, 596 1st Street West, Sonoma, CA 95476, USA}
\altaffiltext{4}{McDonald Observatory, The University of Texas at Austin, Austin, TX 78712, USA}
\altaffiltext{5}{SETI Institute, 189 Bernardo Ave, Suite 100, Mountain View, CA 94043, USA}
\altaffiltext{6}{New York University, Center for Cosmology \& Particle Physics, New York, NY 10003, USA}
\altaffiltext{7}{NASA Senior Fellow}

\setcounter{footnote}{7}

\begin{abstract}
Kepler-91b is a rare example of a transiting hot Jupiter around a red giant star, providing the possibility to study the formation and composition of hot Jupiters under different conditions compared to main-sequence stars. However, the planetary nature of Kepler-91b, which was confirmed using phase-curve variations by Lillo-Box et al., was recently called into question based on a re-analysis of \emph{Kepler} data. We have obtained ground-based radial velocity observations from the Hobby-Eberly Telescope and unambiguously confirm the planetary nature of Kepler-91b by simultaneously modeling the \emph{Kepler} and radial velocity data. The star exhibits temporally correlated noise due to stellar granulation which we model as a Gaussian Process. We hypothesize that it is this noise component that led previous studies to suspect Kepler-91b to be a false positive. Our work confirms the conclusions presented by Lillo-Box et al. that Kepler-91b is a $0.73\pm0.13$ M$_{Jup}$ planet orbiting a red giant star.
%
%
\end{abstract}


\section{Introduction}
The first discovered exoplanets orbiting non-degenerate stars were Jupiter-sized \citep{campbell88} and most orbited just a few stellar radii from their host star \citep{mayor95,marcy96}. This was a milestone moment that demonstrated planetary systems need not resemble our own. While data from the \emph{Kepler} spacecraft have revealed that planetary systems come in many flavors \citep[e.g.][]{lissauer11,carter12,barclay13}, hot Jupiters remain a key area of interest because the large sizes of these planets and their short orbital periods yield the highest signal to noise light curve data with which otherwise undetectable effects can be observed. Examples of this include the detection of in-homogenous clouds \citep{demory13} and the determination of planet masses from Doppler boosting \citep{shporer11,barclay12}.

Additionally the formation of hot Jupiters is still highly debated, with competing theories including migration through the protoplanetary disk \citep{lin96} and high eccentricity migration triggered by dynamical events such as planet-planet scattering \citep{nagasawa08}. The confirmation and characterization of new systems, in particular for evolved stars for which only a handful of hot Jupiters are known, is important to shed light on the origin of these planets.

Kepler-91 was designated KIC 8219268  in the Kepler Input Catalog \citep{brown11}. The star was observed for the entire four year duration of the \emph{Kepler} mission in long cadence mode. The star exhibits stochastic oscillations which confirm that Kepler-91 is an ascending red-giant branch star with a mass of 1.3 $M_\odot$ and radius of 6.3 $R_\odot$ \citep{huber13,lillo14}.


A transiting planet candidate with an orbital period of 6.3 days was detected by the \emph{Kepler} pipeline \citep{jenkins10,tenenbaum13} and assigned Kepler Object of Interest (KOI) number 2133.01 \citep{batalha12}. \citet{lillo14} confirmed the planetary nature of the apparently transiting body. However, the status of this planet has recently been called into question by both \citet{esteves13}, who use phase variations to deduce that the occulting body is self-luminous, and \citet{sliski14} who find that the stellar density derived from a transit model differs significantly from the density calculated using asteroseismic techniques. In this paper we present the results of a light curve model combined with radial velocity observations obtained from the ground and find that the transit-signal is unambiguously caused by a Jupiter-sized planet orbiting the red giant target star\footnote{We reported our radial velocity detection on the Kepler Community Follow-up Observing Program (CFOP) website (\url{http://cfop.ipac.caltech.edu}) in 2012. We encourage the community to make use of this resource.}.

\section{Observational data used in this study}

\subsection{Stellar properties}

The fundamental properties of Kepler-91 have been accurately determined through spectroscopic and asteroseismic analyses using global oscillation properties \citep{huber13} and individual frequency modeling \citep{lillo14} . Both analyses yielded consistent results. Given the increased precision and information on the interior structure when modeling individual frequencies, we have adopted the stellar properties by Lillo-Box et al. in our work (see Table~\ref{tab:stellar}).

\begin{table}
\centering
\caption{Stellar properties adopted from \citet{lillo14}}\label{tab:stellar}
\begin{tabular}{ll}
Property & Adopted value \\
\hline
Effective temperature, $T_{eff}$ (K)		&	$4550\pm75$ \\
Metallically, [Fe/H] (dex)				& 	$0.11 \pm0.07$\\
Mean stellar density, $\rho$ (g cm$^{-3}$)		&	$0.0073 \pm0.0001$ \\
Surface gravity, $\log{g}$ (dex, cgs)		&	$2.953 \pm 0.007$ \\
Stellar mass, $M_\star$ ($M_\odot$)		&	$1.31 \pm 0.10$\\
Stellar radius, $R_\star$ ($R_\odot$)		&	$6.30 \pm 0.16$ \\
\hline
\end{tabular}
\end{table}

\subsection{\emph{Kepler} data}
In this work we utilized the full set of \emph{Kepler} long cadence (29.4-min) observations from the \emph{Kepler} spacecraft obtained over 4 years. These data consist of 17 observational Quarters  (Q1--Q17) where all but the first and last Quarter consist of around 90 days of nearly continuous data. Q1 lasted 40 days and Q17 consists of 31 days of data after which \emph{Kepler} suffered the failure of a reaction wheel.

\begin{figure}
\includegraphics[width=0.47\textwidth]{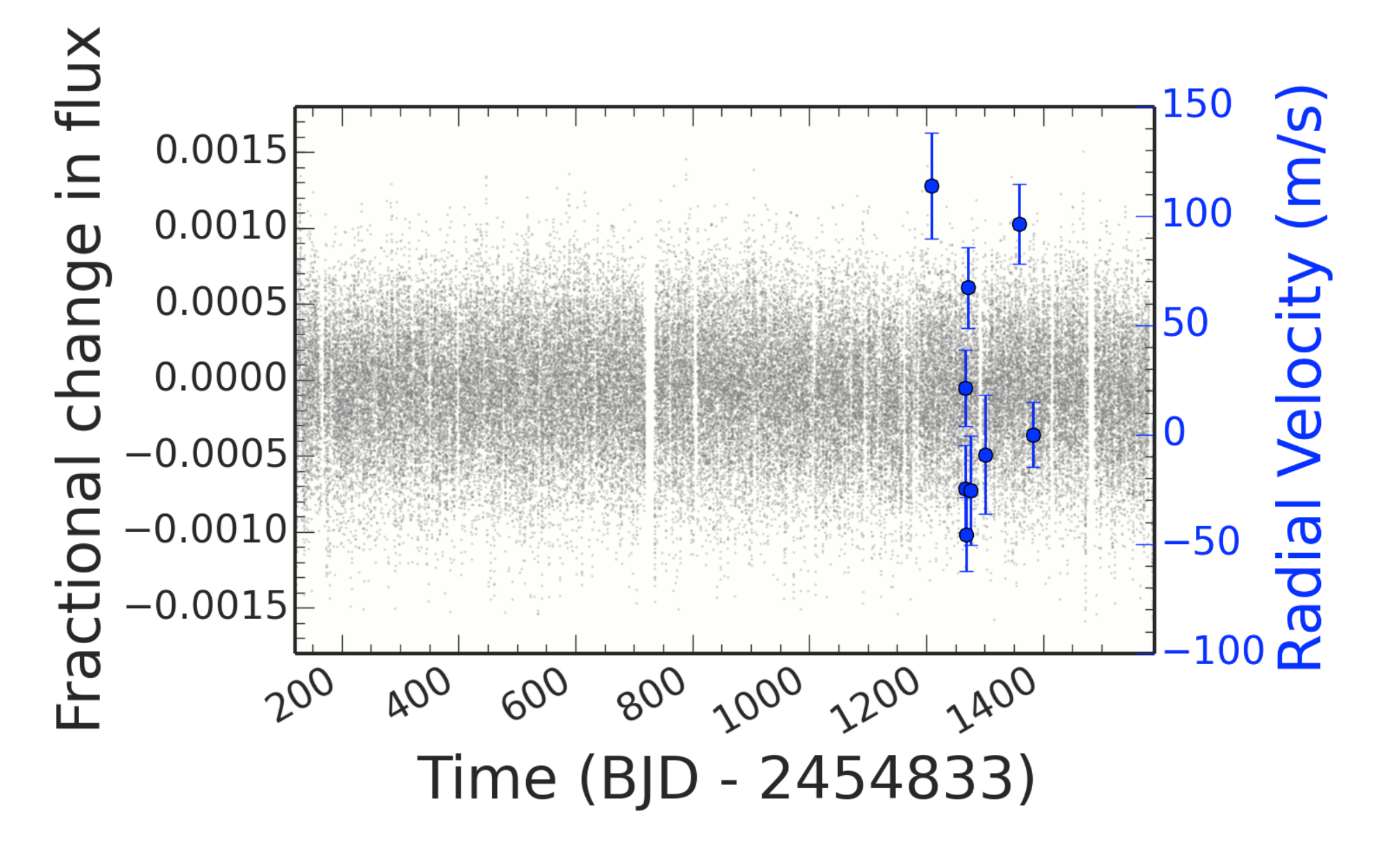}
\caption{The photometric time series we used in the analysis in this work is shown in grey. The light curve has been normalized to zero median. In blue we should the radial velocity data we collected using the Hobby-Eberly Telescope. The photometric and RV data are on the same time axis.}
\label{fig:fulllcrv}
\end{figure}

We used data that have undergone Presearch Data Conditioning \citep{stumpe12,smith12} using the multi-scale maximum a priori (MS-MAP) method \citep{stumpe14}. This preprocessing removes signals related to the spacecraft while retaining much of the variability of astrophysical origin. The MS-MAP algorithm does not entirely retain astrophysical signals, however \citep{DRN21}. While short timescale events such as transits are largely unaffected, signals on timescales of the orbital period of Kepler-91 (6.3 d) are on average damped by about 5\%. The full photometric time series data used in this work in shown in Figure~\ref{fig:fulllcrv}.

\subsection{Radial velocity data}
We obtained precise radial velocity measurements using the High-Resolution-Spectrograph (HRS) \citep{tull98}
instrumental setup for \emph{Kepler} follow-up observations and data reduction algorithms as described in \citet{endl11}. Nine spectra were obtained using an iodine ($I_2$) cell with an exposure time of 900 s and are shown plotted as a function of time in Figure~\ref{fig:fulllcrv}. The data have a resolving power of $R = \lambda/\delta\lambda = 30,000$ and were sky background subtracted. We also obtained two template spectra (without the $I_2$ cell) of Kepler-91, one at $R=30,000$ and a second spectrum with $R=60,000$. The second template was obtained by combining four 900 s exposures and yielded better RV precision. The RV data reported here were obtained using the higher resolution template.

\begin{figure}
\includegraphics[width=0.50\textwidth]{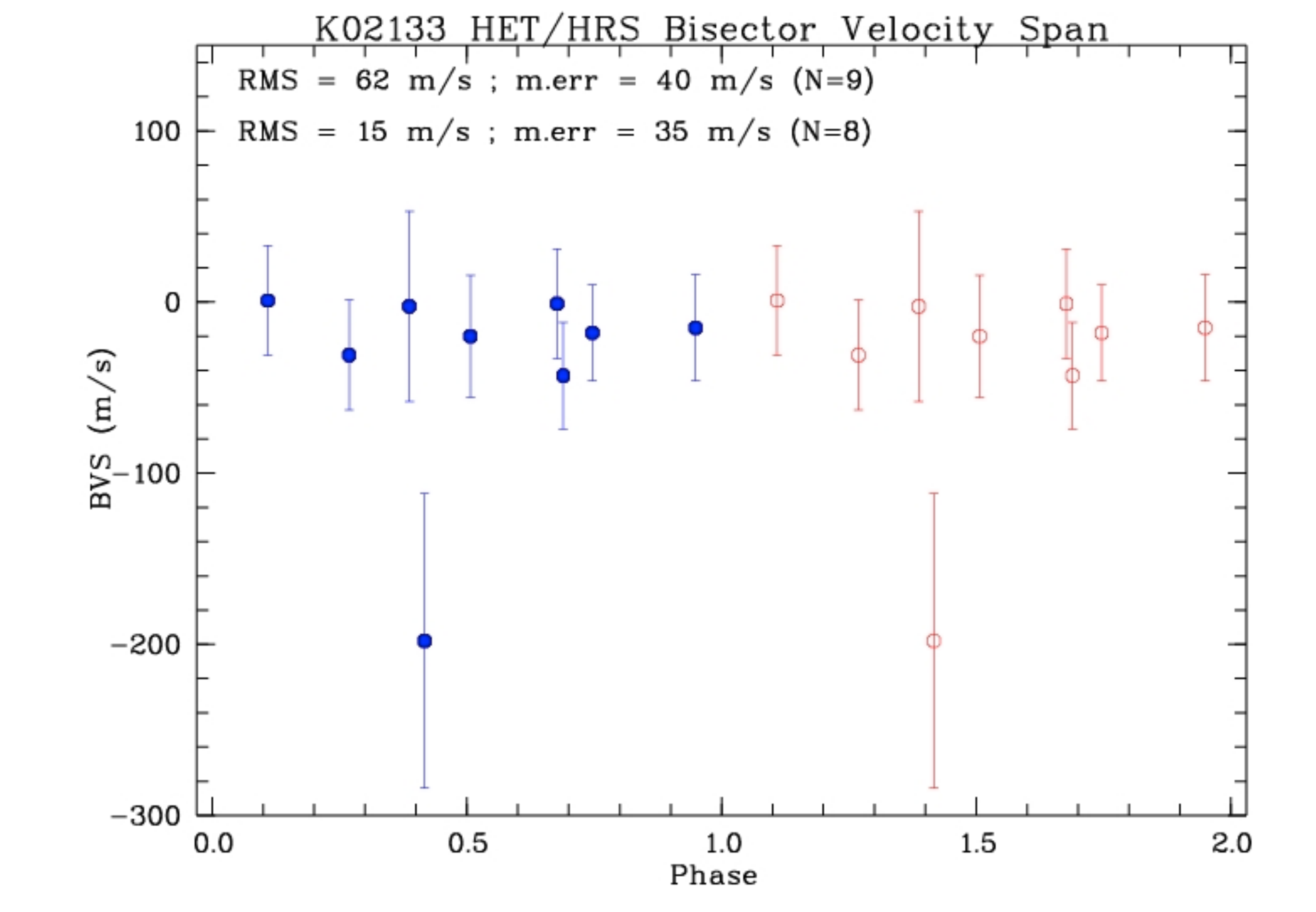}
\caption{The bisector velocity spans of the nine HET observations of Kepler-91. There is a single outlier point which has a very large uncertainty. Excluding this leaves an rms scatter in the bisector velocity spans of 15 m/s. There is not obvious correlation with orbital phase.}
\label{fig:bisectors}
\end{figure}

We also measured bisectors and bisector velocity spans (BVS) for the 9 spectra used for the RV computation. We measured the bisector and BVS of the cross-correlation-function (CCF) for 11 orders that do not contain significant $I_2$ lines. We cross-correlated each spectrum with the $R=30,000$ template to search for variability of the BVS that could indicate a false positive and computed the BVS as the velocity difference of 2 arbitrary points on the CCF bisector at flux values of 0.4 and 0.84, following \citet{hatzes98}. The BVS results are displayed in Figure~\ref{fig:bisectors}. Excluding the poor
quality measurement from the lowest S/N spectrum, the remaining data have a total rms-scatter of 15\,m\,s$^{-1}$ with a mean uncertainty of 35\,m\,s$^{-1}$. The BVS results are consistent with no variability and they also do not correlate with either the orbital phase or the RV measurements.



\begin{table}\label{tab:rv}
\centering
\caption{Radial velocities}
\begin{tabular}{r r r r r}
Time & Velocity\footnote{the values presented here have had an arbitrary offset subtracted to enforce a median of zero}  & Uncertainty & BVS& Uncertainty \\
(BJD-2454833) & (m/s) & (m/s) & (m/s) & (m/s)\\
\hline
1208.86670891		&	114 		&	24		&	-43	&	31\\
1266.71041653		& 	21		&	18		&	-15	&	31\\
1267.70865078		&	-25 		&	20		&	1	&	32\\
1268.70698350		&	-46		&	17		&	-31	&	32\\
1271.68968502		&	67		&	18		&	-18	&	28\\
1275.69264679		&	-26		&	25		&	-3	&	56\\
1300.86443801 	& 	-9		&	27		&	-198\footnote{This outlier comes from the lowest S/N spectrum and is mostly caused by the BVS result from just two of the 11 orders used.}	&	86\\
1358.70858740 	& 	96		&	18		&	-1	&	32\\
1382.63282948 	& 	0		&	15		&	-20	&	36\\
\hline
\end{tabular}
\end{table}


\section{Simultaneous modeling of \emph{Kepler} and RV data}
To provide a self-consistent model of both the light curve and the radial velocity observations we chose to model both datasets simultaneously using the orbital model described in \citet{rowe14}.

Significant planet-induced variability outside of the transit of Kepler-91b has been noted in previous work \citep{lillo14,esteves13}. We chose to model these light curve variations rather than remove them via filtering.  We include 5 physical components in our model of the light curve: a transit, an occultation, ellipsoidal modulation, Doppler boosting and reflection from the planet. We additionally include the radial velocity data as an additional component in the model and finally we include a model for the correlated noise.

\subsection{Parameterization}
We used a limb darkened transit model \citep{mandel02} following a quadratic limb darkening law, and a uniform disk model for the occultation.  The ellipsoidal variations, Doppler boosting and reflection of the planet were modeled in the manner described by \citet{lillo14}, but we parameterized the Doppler beaming in terms of $K$, the radial velocity semi-amplitude, to retain a consistent solution between the light curve data and the spectroscopic radial velocities. The scaling between the radial velocity semi-amplitude and the Doppler beaming amplitude is proportional to a `beaming factor' $B$ such that
\begin{equation}
A_b = B \frac{K}{c} (e\cos{\omega})
\end{equation}
where $A_b$ is the semi-amplitude of the Doppler beaming signal, $c$ is the speed of light, $e$ is the eccentricity and $\omega$ is the argument of periastron. We calculated $B$ in the manner described by \citet{bloemen11} and found a value of 5.46 which we kept fixed. 

We parameterized the combined model in terms of $\rho$ the mean stellar density, $zp$ a photometric zero point nuisance parameter, linear ($\gamma_1$) and quadratic ($\gamma_2$) limb darkening coefficients, $T_0$ the mid-point of transit, $P$ the orbital period of the planet, $b$ the impact parameter, $R_{p}/R_{\star}$ the planet-to-star radius ratio, eccentricity vectors $e\cos{\omega}$ and $e\sin{\omega}$ where $e$ is eccentricity and $\omega$ is the argument of periastron, the amplitude of the ellipsoid variations $A_e$, the amplitude of the reflection from the planet $A_r$, the occultation depth $F_e$, radial velocity semi-amplitude $K$, and $V$ a radial velocity zero-point. We account for the 29.4-min integration time of the exposures by subsampling the model 11 times per observation then integrating over these subsamples \citep{kipping10}.

Because we are using disparate data sets is it useful to include a parameter for both the light curve and radial velocity data that is an additional noise term which is added in quadrature with the formal uncertainty ($\sigma_{lc}$ and $\sigma_{rv}$). These account for missing physics on our model in addition to dealing with underestimation of the reported uncertainties. This creates a flexible model that enables us to scale the two data sets appropriately. We sample in the natural log of $\sigma_{lc}$ an $\sigma_{rv}$ which is equivalent to using a Jeffery's prior on these parameters.

\subsection{Gaussian Process noise model}
Red giants are known to show significant correlated noise on timescales of hours to weeks due to stellar granulation (Mathur et al. 2012), which is also clearly apparent in Kepler-91. Not accounting for this correlated noise component can bias the observed planet parameters \citep{carter09}.

A Gaussian process is a general framework for modeling correlated noise
\citep{rasmussen} and it has been applied to transmission spectroscopy
\citep{gibson-gp,evans-gp} and \emph{Kepler} light curves \citep[][Dawson
\emph{et al.}\ in press]{ambikasaran14}.
The basic idea is to model the light curve as being drawn from a large
$N_\mathrm{data}$-dimensional Gaussian where the covariance matrix $K$ is
modeled as
\begin{eqnarray}
K_{ij} &=& \sigma_i^2\,\delta_{ij} + k(t_i,\,t_j)
\end{eqnarray}
where $\sigma_i$ is the observational uncertainty, $\delta_{ij}$ is the
Kronecker delta, and $k(t_i,\,t_j)$ is the so-called \emph{covariance kernel
function} that quantifies the correlations between data points.
There is a lot of freedom in how we choose this covariance function but here
we choose a very simple and commonly used model called the
squared-exponential or radial basis function kernel
\begin{eqnarray}
k(t_i,\,t_j) &=& A^2\,\exp\left(-\frac{[t_i-t_j]^2}{2\,\tau^2}\right)
\end{eqnarray}
where the covariance amplitude $A_{GP}$ is measured in flux units and the length
scale $\tau_{GP}$ is measured in days. We chose the squared-exponential kernel because it exhibits a number of traits that make it suitable for modeling granulation in red giant light curves: (a) it is smooth with no sharp discontinuities, and (b) the covariance between observed points is only a function of the distance (in time) between points.
When performing our analysis, we first subtract a realization of the physical light curve model, then calculate the likelihood of the residuals given the covariance matrix defined by the Gaussian Process model. We marginalize over the parameters $A_{GP}$ and $\tau_{GP}$ with uniform priors in our MCMC analysis.

\subsection{Priors}
Priors on our model parameters are shown in Table~\ref{tab:priors}. Of particular note is $\rho$, where we enforce a normal prior constrained from the probability density found from asteroseismology. We enforce a $1/e$ prior because without this our parameterization in terms of $e\sin{\omega}$ and $e\cos{\omega}$ would bias $e$ high \citep{eastman13}. We use a normal prior on limb darkening with the expectation obtained through interpolation of model limb darkening in the \emph{Kepler} bandpass with $T_{eff}$, $\log{g}$ and [Fe/H] fixed at the values shown in Table~\ref{tab:priors}, and with a standard deviation of 0.1. Finally, we set a number of prior constraints on linear combinations of $\gamma_1$ and $\gamma_2$ that prevent them from taking unphysical values \citep{burke08}.

\begin{table}
\centering
\caption{Model parameters}\label{tab:priors}
\begin{tabular}{l l }
Property & Prior \\
\hline
$\rho$ (g/cc)		&	$\mathcal{N}(0.0073;0.0001)$ \\
$zp$				& 	$\mathcal{U}($-$0.0002;0.0002)$\\
$\gamma_1$		&	$\mathcal{N}(0.67;0.6)$ \\
$\gamma_2$		&	$\mathcal{N}(0.09;0.6)$\\
$T_0$ (BKJD)& $\mathcal{U}(136.3;136.5)$\\
$P$ (days) & $\mathcal{U}(6.2;6.3)$\\
$b$ & $\mathcal{U}(0;1+R_{p}/R_{\star})$\\
$R_{p}/R_{\star}$ & $\mathcal{U}(0;1)$ \\
$e\cos{\omega}$ &$\mathcal{U}(-1;1)$ \\
$e\cos{\omega}$ & $\mathcal{U}(-1;1)$\\
$e$ & $1/e$\\
$A_e$ (ppm)&$\mathcal{U}(-500;500)$ \\
$A_r$ (ppm)& $\mathcal{U}(-500;500)$\\
$F_e$ (ppm)& $\mathcal{U}(-500;500)$\\
$V$ (m/s)&$\mathcal{U}(-300;300)$ \\
$K$ (m/s)&$\mathcal{U}(-1000;1000)$ \\
$\log \sigma_{lc}$ & $\mathcal{U}(-50;0.0)$\\
$\log \sigma_{rv}$ (m/s)&$\mathcal{U}(-15;6)$ \\
$A_{GP}$ &  $\mathcal{U}(0;0.001$ \\
$\tau_{GP}$ &  $\mathcal{U}(0;0.5)$ \\
\hline
\end{tabular}
\end{table}

\subsection{Markov-Chain Monte Carlo modeling}
We numerically integrated the posterior probability using an efficient affine invariant Markov-Chain Monte Carlo (MCMC) algorithm \citep{goodman10,foreman13}.  This method utilizes many walkers to reduce autocorrelation time; we opted to use 700 walkers each taking 15,000 steps for a total of $10.5\times10^6$ samples of the posterior probability. However, we discard the first 5,000 samples in each walker as burn-in which leaves $7\times10^6$ samples used to calculate posterior distributions.

\begin{figure}
\includegraphics[width=0.50\textwidth]{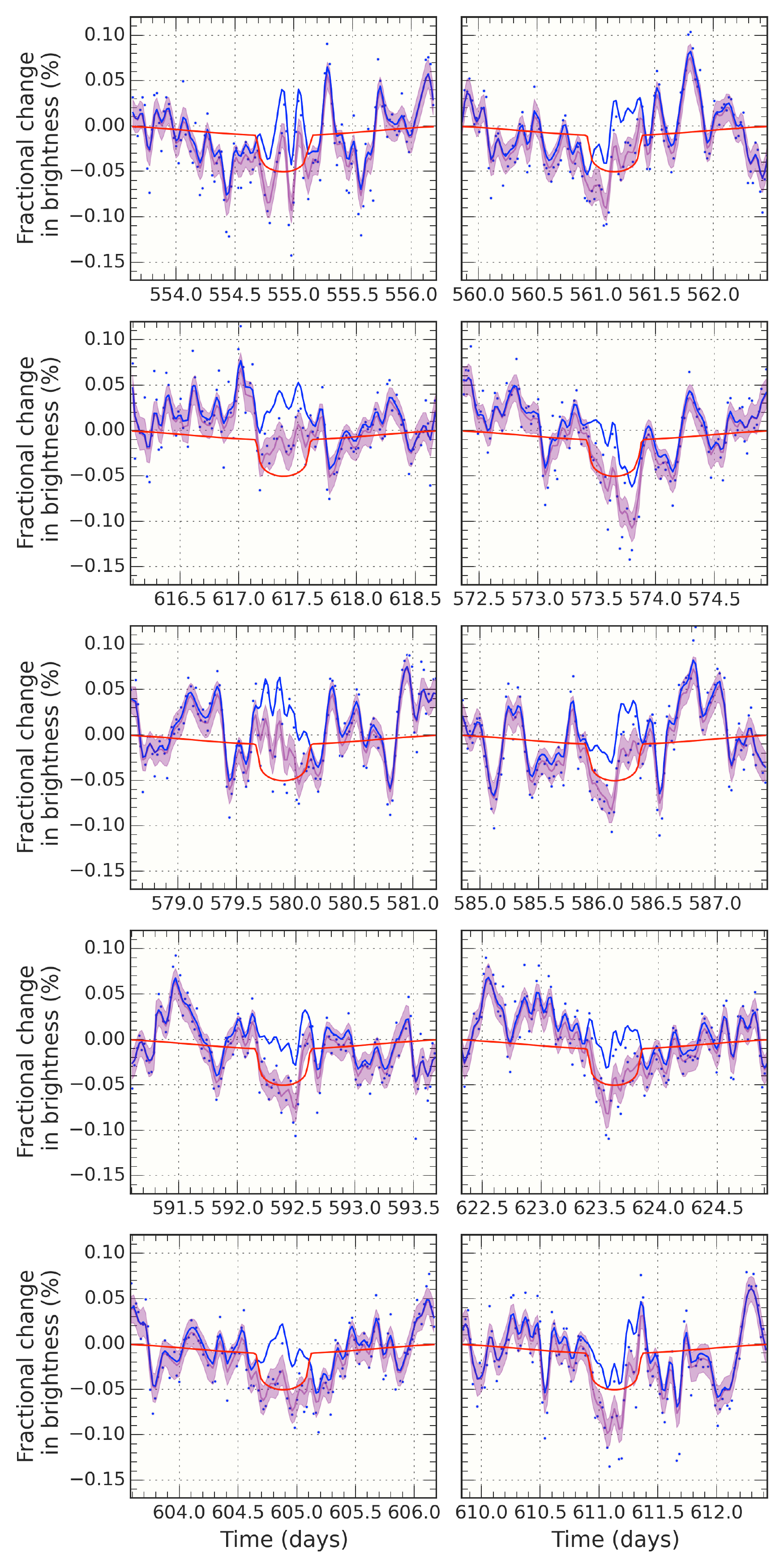}
\caption{Ten transits are shown to demonstrate our noise model. The observed data is shown as black points, the mean noise model in blue and the mean light curve model in red. The combined light curve and noise model is shown as the shaded purple region where the shaded region shows the 1-$\sigma$  bounds of the model. The shaded region does not include the white noise component but does include the uncertainty in both the noise model and the transit model. For each transit we show 2.4 days of data, while the y-scale is 0.29\%.}
\label{fig:filter}
\end{figure}

In Figure~\ref{fig:filter} we show ten transits seen in the \emph{Kepler} data, the mean noise model, the light curve model and the 1-$\sigma$ uncertainty on the combined light curve and noise model. The combination of our noise model and transit model does a good job of matching the correlated noise and planet signals seen in the data.

\section{Results}
We were able to produce a self consistent model that described the data well. Parameters from the modeling are reported in Table~\ref{tab:results}, we give the median and central 68\% bounds of the marginalized posterior distribution for each parameter. We also list the sample that -- in our chains -- obtained the highest probability. This so-called ``best fit'' is \emph{not} the result of an optimization but it does have a self-consistent set of parameters and if you need a single model, this is the one to use. However, if one needs to draw conclusions about the physical properties of the planet then we advise using the median with reported uncertainty.

\begin{figure*}
\includegraphics[width=0.97\textwidth]{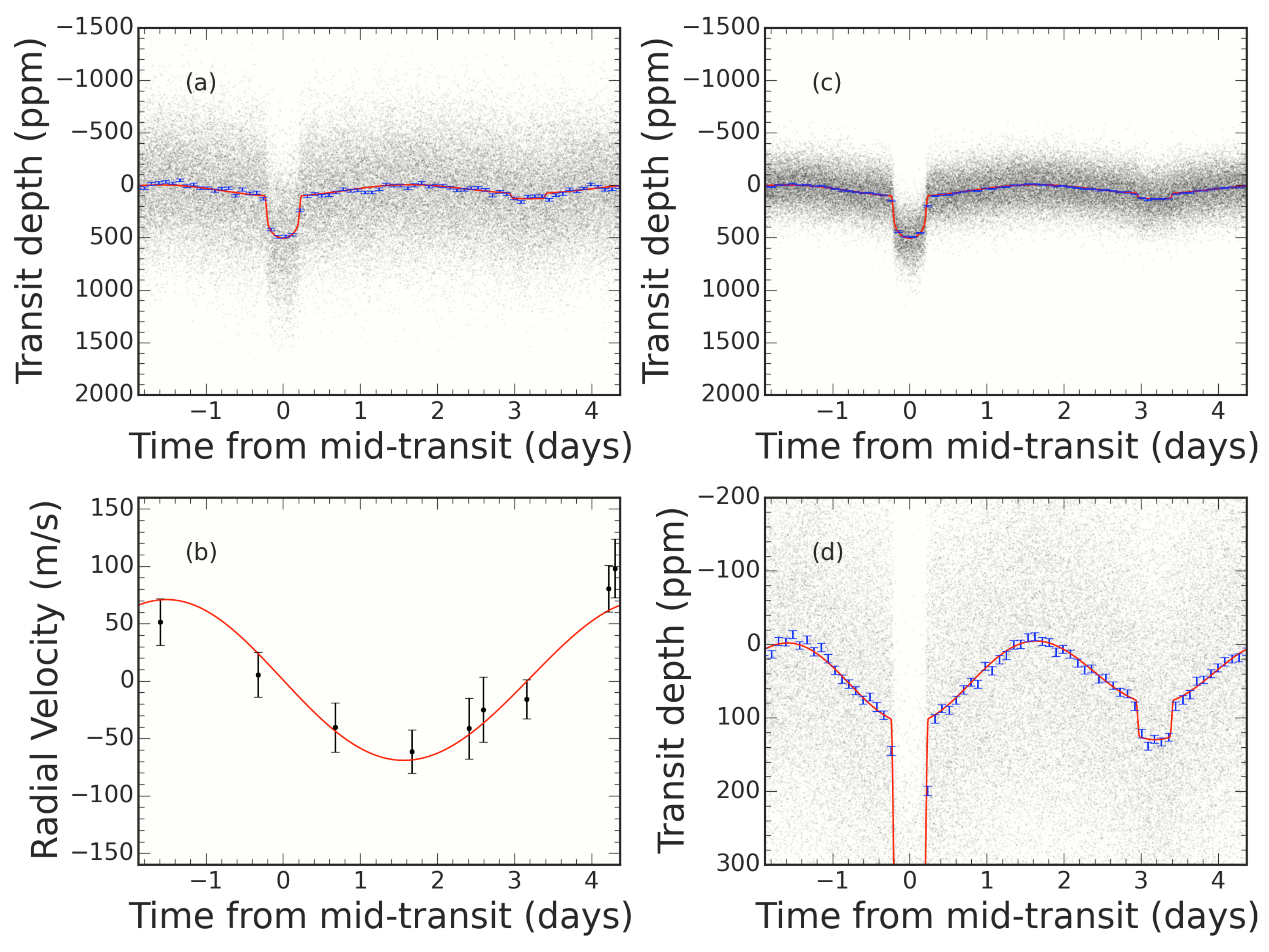}
\caption{Observed data is shown in black, the best fitting model in red and binned data is shown in blue. The data have been folded on the orbital period of the planet. Panel (a) shows the observed data as black semi-transparent points, the large scatter is due to correlated noise in the observed data which has not been modeled in this plot. The blue points shown are binned observed data with 1000 observed points included per bin. The red curve is the best fitting light curve model, excluding the noise model. Panel (b) shows our observed radial velocity data in black and the best fitting model in red folded on the same phase as the Kepler data. Panel (c) is the same as panel (a) but the mean Gaussian Process noise model has been subtracted from the observed data. The scatter on the data is the white noise component of the noise plus any remaining correlated noise we have not modeled. Panel (d) is a zoom in on panel (c) to show the phase variations from Kepler-91b and the secondary eclipse.}
\label{fig:results}
\end{figure*}

In Figure~\ref{fig:results} we show, in the upper left panel the observed data in black (this data still contains correlated noise) with the best-fit model in red and binned observed data in blue. The data have been folded on the orbital period of the planet. The lower left shows the RV data folded on the orbital period of the planet with the same phase as the top right panel. The upper right panel shows the data with a mean Gaussian Process noise model subtracted. Subtracting the mean model gives an idea of what the Gaussian Process is modeling and what noise level remains in the data. We stress that the mean noise model subtracted data should never be used for inference and is for demonstration purposes only. The lower right panel is the same as the upper right panel but zoomed in to better show the ellipsoidal variations and the secondary eclipse.

\begin{figure}
\includegraphics[width=0.50\textwidth]{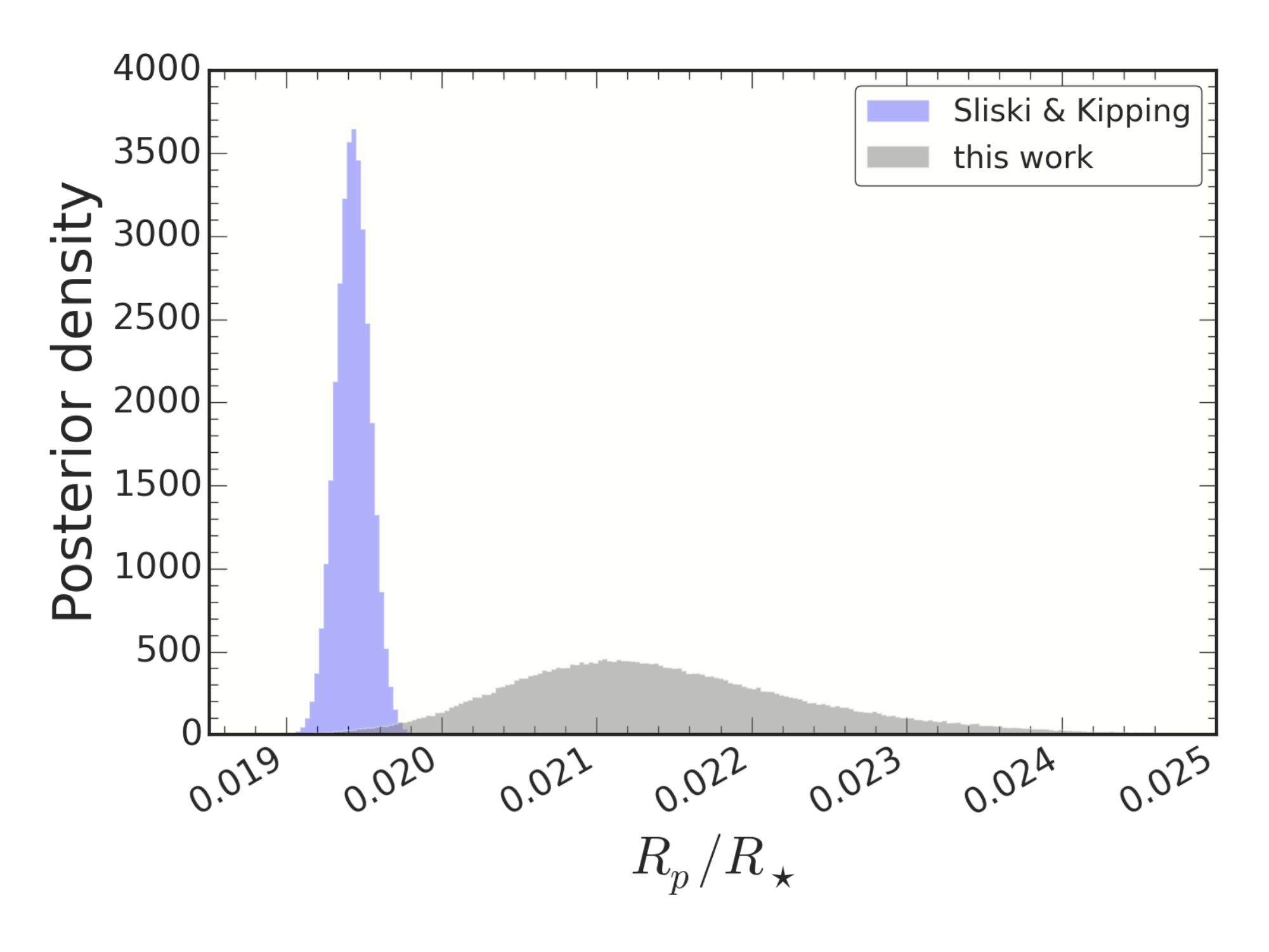}
\caption{Our marginal distribution on the model parameters $R_{p}/R_{\star}$ compared with \citet{sliski14}. Our estimate has a much larger uncertainty. The estimate from this paper differs with the estimate of \citet{sliski14} by about 2.5 standard deviations.}
\label{fig:rprs}
\end{figure}

Our parameter estimates are largely consistent with \citet{lillo14}. While we obtain a slightly lower estimate for $R_{p}/R_{\star}$, we are still consistent with \citet{lillo14} at the level of $<$1.5$\sigma$. Our estimate of $R_{p}/R_{\star}$ is inconsistent with that found by \citet{sliski14}, as shown in Figure~\ref{fig:rprs}. The probability that our $R_{p}/R_{\star}$ estimate agrees with the value found by \citet{sliski14} is 1\%.

We derive a planet mass of $0.73\pm 0.13$ M$_{Jup}$ which, combined with a planet radius of $1.308^{+0.061}_{-0.074}$ $R_{Jup}$ yields a density of $0.40^{+0.10}_{-0.09}$ g cm$^{3}$ implying that Kepler-91b is somewhat inflated \citep{baraffe10}. The consistency between the planet mass we obtain with radial velocity observations and the mass \citet{lillo14} obtain from the phase curve alone is remarkably good (\citeauthor{lillo14} determined a mass for Kepler-91b of $0.88^{+0.17}_{-0.33}$ M$_{Jup}$). Kepler-91b is one of a growing number of examples demonstrating that hot Jupiter densities can be estimated from \emph{Kepler} data alone \citep[e.g.][]{barclay12,quintana13,esteves13}.

\begin{figure}
\includegraphics[width=0.50\textwidth]{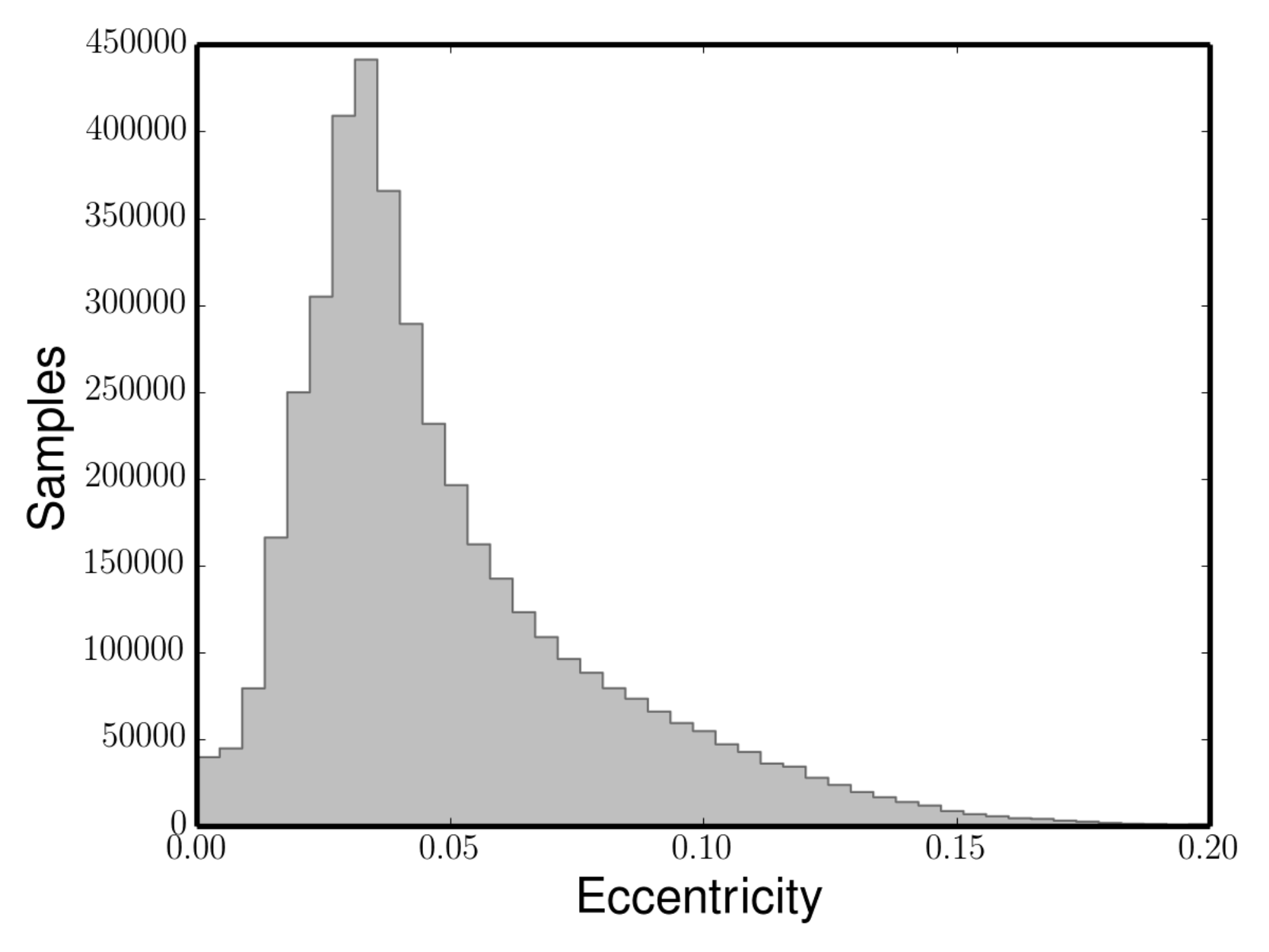}
\caption{The posterior distribution for orbital eccentricity for Kepler-91b. While a slightly eccentric orbit is preferred by the data, we cannot rule out a circular orbit model.}
\label{fig:ecc}
\end{figure}

\citet{lillo14} suggest that Kepler-91b may be on an eccentric orbit. In Figure~\ref{fig:ecc} we show our posterior distribution for eccentricity derived from our $e\sin{\omega}$ and $e\cos{\omega}$ samples. While the data prefers an eccentric model, a circular orbit cannot be ruled out.

We note that, as mentioned in Section~2, signals on timescales of more than a few days are damped by a few percent owing to the detrending performed within the Kepler pipeline. The only low frequency signal that we significantly detected in these data was from stellar ellipsoidal variations. With an uncertainty of 10\% on the amplitude of the ellipsoidal signal, a 5\% damping of the signal would not significantly change our results. It should be kept in mind that signals with longer timescales (more than 15--20 days) are significantly damped by the detrending method used within the Kepler pipeline.

The additional uncertainty we report on the radial velocity observations, $\sigma_{rv}$, accounts for astrophysical noise sources that we do not include in our model. One of the sources of astrophysical noise that $\sigma_{rv}$ accounts for is the radial velocity \emph{jitter} induced by the stellar oscillations. The oscillation signal expected is a few meters per second \citep{kjeldsen95} and given its short timescale relative to the planetary orbital period  acts as an additional noise source on the data. However, photon noise is the still dominant source of uncertainty.

\begin{table}
\caption{Summary of posterior probabilities from the MCMC modeling. Parameters below the line break are derived from the parameters sampled by the model}\label{tab:results}
\begin{tabular}{l l l l l}
Parameter&Best fit& Median&84.1\%&15.9\%\\
\hline
%
$\rho$ (g/cc)			&	0.00724		&	0.00730		&	+0.00010		&	$-$0.00010	\\
$zp$	(ppm)			&	$-$56.8		&	$-$60.0		&	+8.0			&	$-$4.7		\\
$\gamma_1$			&	0.43			&	0.63			&	+0.39		&	$-$0.43		\\
$\gamma_2$			&	0.27			&	0.12			&	+0.46		&	$-$0.48		\\
$T_0$ (BKJD\footnote{BKJD is the time system used by \emph{Kepler} and is defined by Barycentric Julian Date (BJD) $-$ 2454833}) 	&	136.3837		&	136.3846		&	+0.0042		&	$-$0.00427	\\
$P$ 	(days)			&	6.246696		&	6.246713		&	+0.000040	&	$-$0.000032	\\
$b$ 					&	0.865		&	0.858		&	+0.014		&	$-$0.016		\\
$R_{p}/R_{\star}$		&	0.0212		&	0.0211		&	+0.0014		&	$-$0.0008		\\
$e\cos{\omega}$ 		&	0.0134		&	0.0218		&	+0.0080		&	$-$0.00860	\\
$e\sin{\omega}$ 		&	-0.012		&	-0.016		&	+0.022		&	$-$0.028		\\
$A_e$ (ppm)			&	50.8			&	52.3			&	+5.0			&	$-$4.9		\\
$A_r$ (ppm)			&	27			&	32			&	+10			&	$-$19		\\
$F_e$ (ppm)			&	45			&	40			&	+19			&	$-$18		\\
$V$ 	(m/s)				&	15.6			&	15.9			&	+6.1			&	$-$6.3		\\
$K$ 	(m/s)				&	70.2			&	67.1			&	+9.4			&	$-$8.3		\\
$\log \sigma_{lc}$ (ppm)	&	-15.3			&	-15.9			&	+2.0			&	$-$1.7		\\
$\log \sigma_{rv}$ (m/s)	&	3.0			&	2.3			&	+1.9			&	$-$1.7		\\
$A_{GP}$ (ppm)		&	301.6		&	301.2		&	+1.8			&	$-$1.9		\\
$\tau_{GP}$  (days)		&	0.03717		&	0.03712		&	+0.00027		&	$-$0.00031	\\
%
%
%
\hline
%
$R_{p}$ ($R_{Jup}$) 	&	1.300		&	1.308		&	+0.061		&	$-$0.074		\\
$M_{p}$ ($M_{Jup}$)	&	0.76			&	0.73			&	+0.13		&	$-$0.13		\\
$a/R_{\star}$			&	2.463		&	2.469		&	+0.011		&	$-$0.011		\\
$A_g$				&	0.66			&	0.52			&	+0.22		&	$-$0.20		\\
$e$					&	0.018		&	0.040		&	+0.040		&	$-$0.016		\\
$i$ (deg)				&	69.17		&	69.12		&	+0.58		&	$-$0.88		\\
$\rho_{p}$ (g/cc) 		&	0.43			&	0.40			&	+0.10		&	$-$0.09		\\
\hline
\end{tabular}
\end{table}

\section{Discussion}
\begin{figure}
\includegraphics[width=0.50\textwidth]{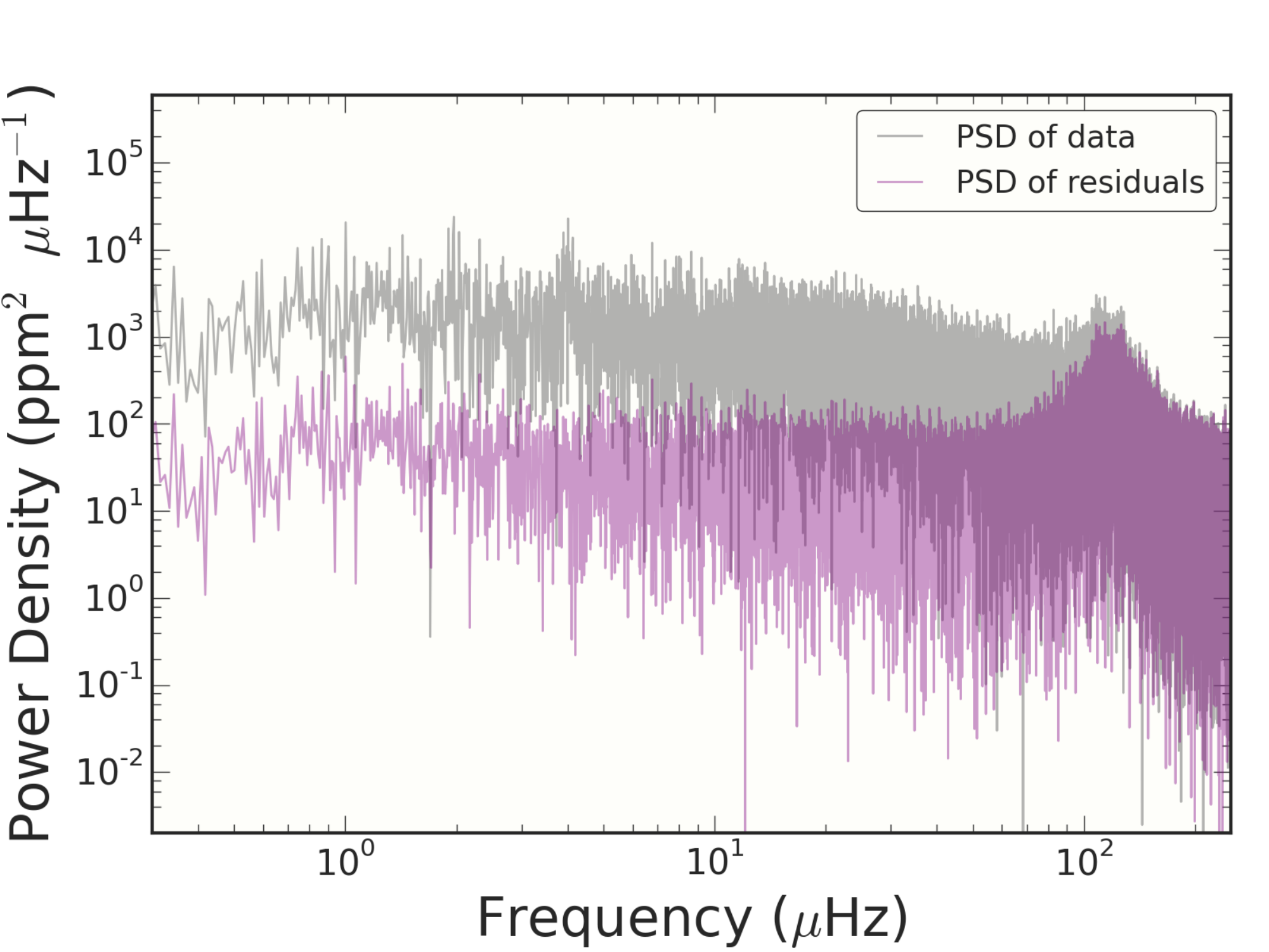}
\caption{The power spectral density of the light curve of Kepler-91 is shown in grey while the same light curve with our mean noise model and best fitting transiting planet model subtracted off is shown in purple. The low frequency power is reduced by an order of magnitude by using our noise model. The peak around 100 $\mu$Hz is owing to the red giant oscillations which we make no attempt to remove.}
\label{fig:PSD}
\end{figure}

\begin{figure}
\includegraphics[width=0.50\textwidth]{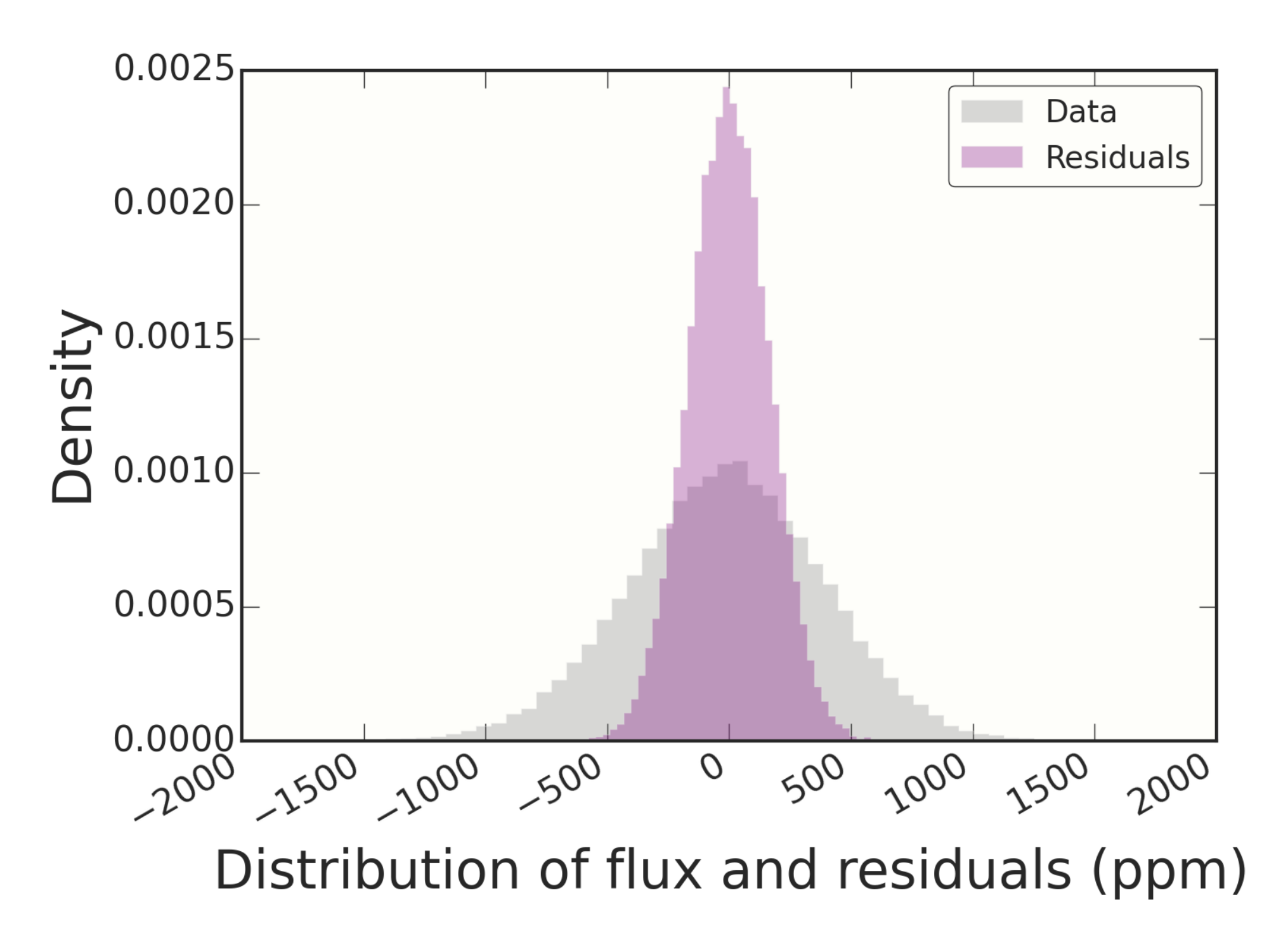}
\caption{Histograms of the values of the light time series of Kepler-91. In grey is the flux time series as extracted from the observed data. In purple are the residuals of the flux time series with a mean noise model removed. The residuals have a much lower scatter than the original data. The standard deviation of the original data is 398 ppm while the standard deviation of the residuals is 166 ppm.}
\label{fig:hist_resid}
\end{figure}

We obtained good fits to both the radial velocity and light curve data with our model as shown in Figure~\ref{fig:results}. Our Gaussian process model is intended to describe the surface granulation on the red giant. In Figure~\ref{fig:PSD} we show the power spectral density of the original light curve in grey and in purple we show the power spectral density for the light curve with the mean noise model and best fitting transit model subtracted out. We can clearly see that the low frequency noise is well fit by the model. We note that the red giant oscillations are not removed by this model and can be seen as a Lorentzian peak around 100 $\mu$Hz. In Figure~\ref{fig:hist_resid} we show the histogram of our flux values from the light curve (with median of zero) and the residuals of flux time series with the mean nose model and transiting planet model subtracted. It is clear here that the residuals with the noise model removed are significantly smaller than in the standard deviation of original time series.

The radial velocity observations phase well with the orbital period defined by the transit and therefore almost certainly are caused by a reflex motion of the star that the planet orbits. The transit model fits the data well with low eccentricity. If the density constraint from asteroseismology was found to be a poor choice of prior we would need a model with high eccentricity, which is not seen. The mass we derive is consistent with a planetary body. We therefore conclude Kepler-91b is a \emph{bona fide} planet.

The validation presented in this work raises questions as to why Kepler-91 was previously suspected to be a false-positive. \citet{esteves13} concluded that Kepler-91b is a false positive based on a phase curve fit where they find a night-side temperature inconsistent with a body that is not self luminous. We do not draw the same conclusions. Using the same equation as \citeauthor{esteves13} for geometric albedo, we derived a value of $0.52^{+0.22}_{-0.20}$ compared with $2.49^{+0.55}_{-0.60}$ from \citeauthor{esteves13}. 

When modeling the correlated noise we find a transit depth that is inconsistent with the transit depth found by both \citet{esteves13} and \citet{sliski14} who both report shallower transits. The method we use to calculate the geometric albedo is
\begin{equation}
A_g = F_e \left(\frac{a}{R_p}\right)^{2} = F_e \left(\frac{a}{R_{\star}}\frac{R_{\star}}{R_{p}}\right)^{2} .
\end{equation}
Therefore, underestimating $R_{p}/R_{\star}$ will overestimate the albedo $A_g$ and lead to a conclusion that the occulting body is self-luminous.

\begin{figure}
\includegraphics[width=0.50\textwidth]{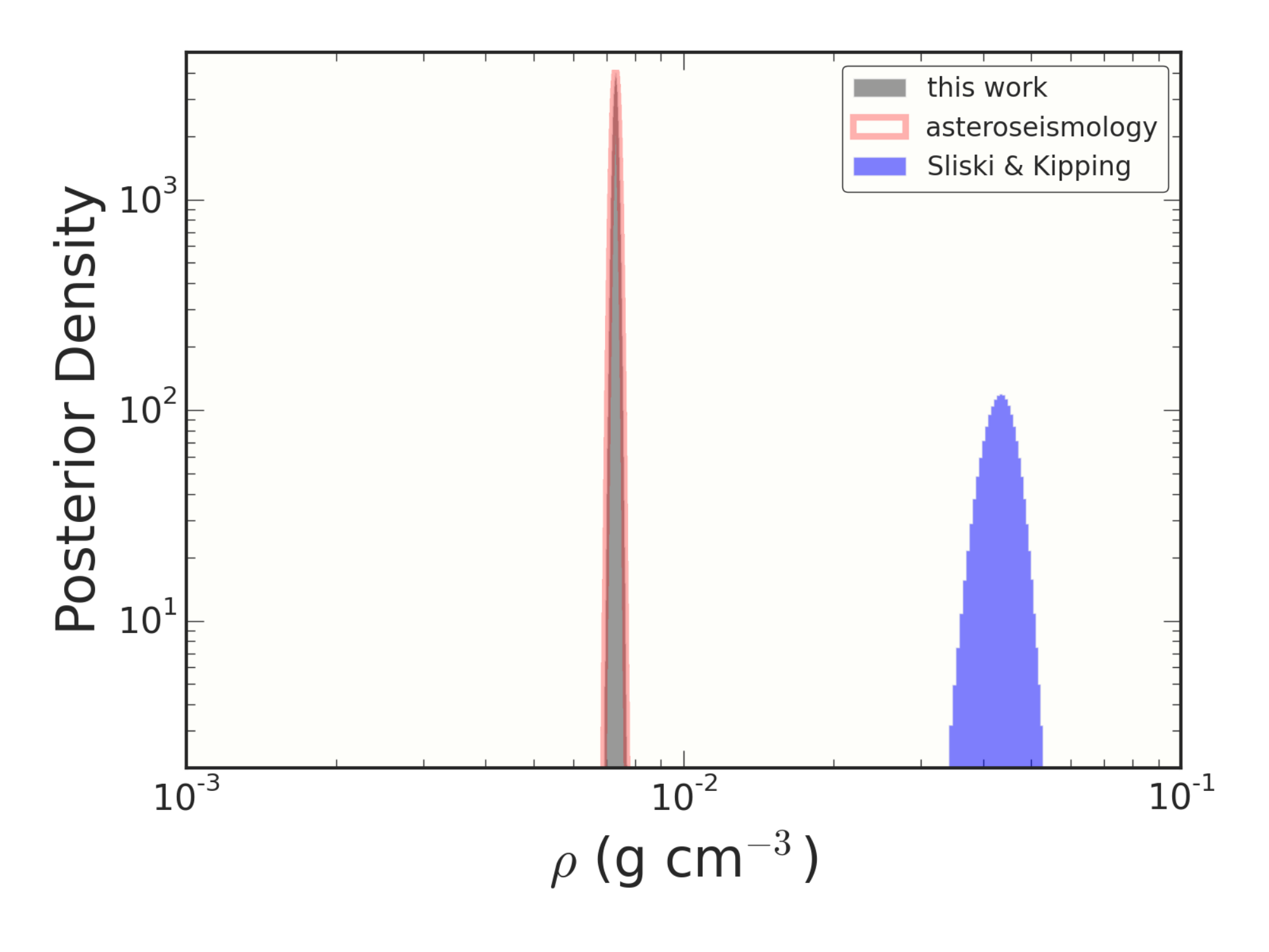}
\caption{Our marginal distribution on the model parameters $\rho$ compared with \citet{sliski14}. The two estimates differ very significantly. The value from asteroseismology, which we use as a prior, is shown in red.}
\label{fig:rho}
\end{figure}

\citet{sliski14} find that the mean stellar density derived purely from a transit model is not consistent with the asteroseismic density of the star. In Figure~\ref{fig:rho} we show our estimate of the stellar density and the estimate of \citet{sliski14}, the two estimates differ very significantly. They hypothesize that the observed transit is not caused by a body occulting the target star which is inconsistent with our results.

\begin{figure}
\includegraphics[width=0.50\textwidth]{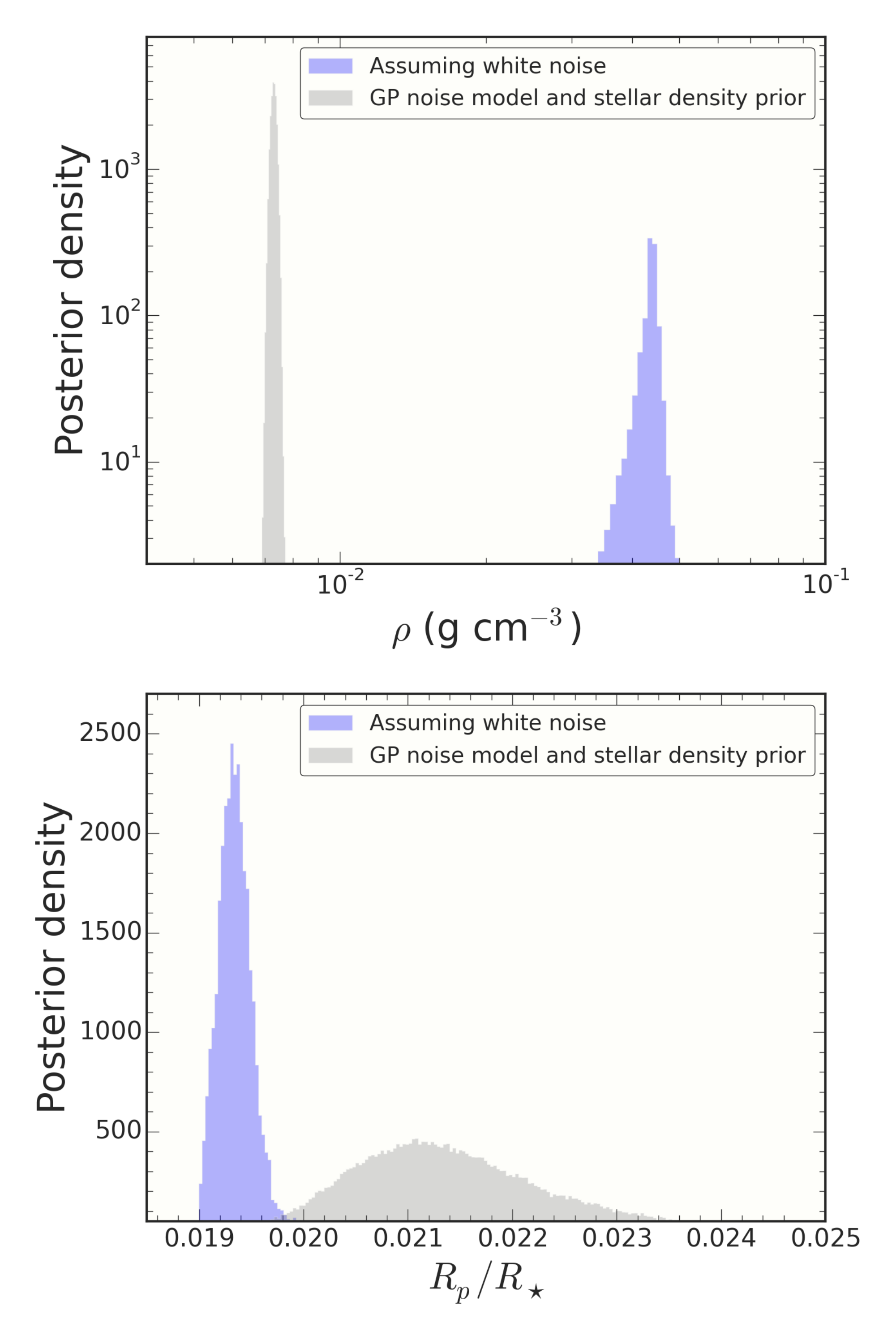}
\caption{The two panels show a comparison of our posterior model parameters when using the Gaussian Process noise model and a prior on the stellar density from the asteroseismic constraint with the posterior when assuming no time correlated noise. The upper panel shows the mean stellar density, $\rho$ while the lower panel shows the planet-to-star radius ratio $R_p/R_\star$. Our posteriors when not including the correlated noise mirrors the results of \citet{sliski14} but when we include the noise model our results are consistent with the conclusion we must draw from the radial velocity data that Kepler-91b is a planet.}
\label{fig:comp_noGP}
\end{figure}

We have hypothesized that the different conclusions regarding the planetary nature of Kepler-91b are owing to correlated noise in the Kepler data. With this in mind, we performed an additional MCMC simulation but this time did not include our Gaussian Process noise model and removed the asteroseismic prior on the stellar density. We found that without the Gaussian Process model our posteriors mirrored the results found by \citet{sliski14} with $\rho$~=~$0.043^{+0.001}_{-0.003}$ g cm$^{-3}$ and $R_p/R_\star$~=~$0.01955\pm 0.0002$, in tension to our original parameters when using the noise model (a comparison is shown in Figure~\ref{fig:comp_noGP}). This test shows that it is the Gaussian Process noise model that cause the differing results from \citet{sliski14}. 

We believe it is reasonable to draw the conclusion that not including correlated noise in the model for this star is what led to the erroneous results of the aforementioned teams. However, it is important to note that it is only because we were in possession of the radial velocity data, which raised doubts on the interpretation of \citeauthor{esteves13} and \citeauthor{sliski14}, that we considered the more sophisticated noise model described here.

\section{Conclusions}
We combined \emph{Kepler} data presented in previous studies of Kepler-91 with radial velocity data obtained from the Hobby-Eberly Telescope. We find that these data can be fit in a self consistent manner and the results lead us to conclude that Kepler-91b is a planet, validating the work of \citet{lillo14}. We hypothesize that recent work claiming this as a false positive erred owing to the challenge of modeling a star with a strong correlated noise component and relatively high amplitude out-of-transit variations. We use a Gaussian Process to model the correlated noise present in the light curve of this red giant star and we derive parameters that are in agreement with the planetary nature of Kepler-91b. We suggest that Gaussian Processes may prove to be very useful in future studies of transiting exoplanets orbiting stars exhibiting correlated noise properties.

The validation of Kepler-91 presented in this work confirms that hot Jupiters exist in close-in orbits around red giant stars. Future studies of Kepler-91 and similar planets will allow insights on the internal composition of hot Jupiters as a function of host star evolution \citep{spiegel12}, and measurements of the spin-orbit inclination of such systems may allow us to test theories of obliquity damping that have been proposed for hot Jupiter populations around less evolved stars \citep{winn10}.


\emph{We note that after the submission of our paper, independent radial velocity observations confirming the planetary nature of Kepler-91b were published by \citet{lillo14b}.}

\acknowledgments
This paper includes data collected by the \emph{Kepler} mission. Funding for the \emph{Kepler} mission is provided by the NASA Science Mission Directorate. We would like to express our gratitude to all those who have worked on the \emph{Kepler} pipeline over the many years of the \emph{Kepler} mission. Some \emph{Kepler} data presented in this paper were obtained from the Mikulski Archive for Space Telescopes (MAST) at the Space Telescope Science Institute (STScI). STScI is operated by the Association of Universities for Research in Astronomy, Inc., under NASA contract NAS5-26555. Support for MAST for non-HST data is provided by the NASA Office of Space Science via grant NNX09AF08G and by other grants and contracts. Our MCMC sampling was performed on the Pleiades supercomputer of the NASA Advanced Supercomputing Division at NASA's Ames Research Center. We used data obtained from The Hobby-Eberly Telescope (HET), a joint project of the University of Texas at Austin, the Pennsylvania State University, Stanford University, Ludwig-Maximilians-Universit\"{a}t M\"{u}nchen, and Georg-August-Universit\"{a}t G\"{o}ttingen. The HET is named in honor of its principal benefactors, William P. Hobby and Robert E. Eberly. We thank the Statistical and Applied Mathematical Sciences Institute (SAMSI) for hosting the Modern Statistical and Computational Methods for Analysis of Kepler Data workshop where the idea for this work came about. The work performed at SAMSI was partially supported by the National Science Foundation under Grant DMS-1127914 to the Statistical and Applied Mathematical Sciences Institute. Any opinions, findings, and conclusions or recommendations expressed in this material are those of the author(s) and do not necessarily reflect the views of the National Science Foundation. We thank Ruth Angus (Univ. of Oxford) for valuable discussions on noise sources in RV observations. We thank Jeffrey C. Smith, Joe Catanzarite (both SETI Inst.) and David Kipping (Harvard/CfA) for suggestions on how to improve the manuscript. E.V. Quintana is supported by a NASA Senior Fellowship at the Ames Research Center, administered by Oak Ridge Associated Universities through a contract with NASA. D. Huber acknowledges support by NASA under Grant NNX14AB92G issued through the Kepler Participating Scientist Program.




\begin{appendix}
\section{Additional plots}
\begin{figure}
\includegraphics[width=0.97\textwidth]{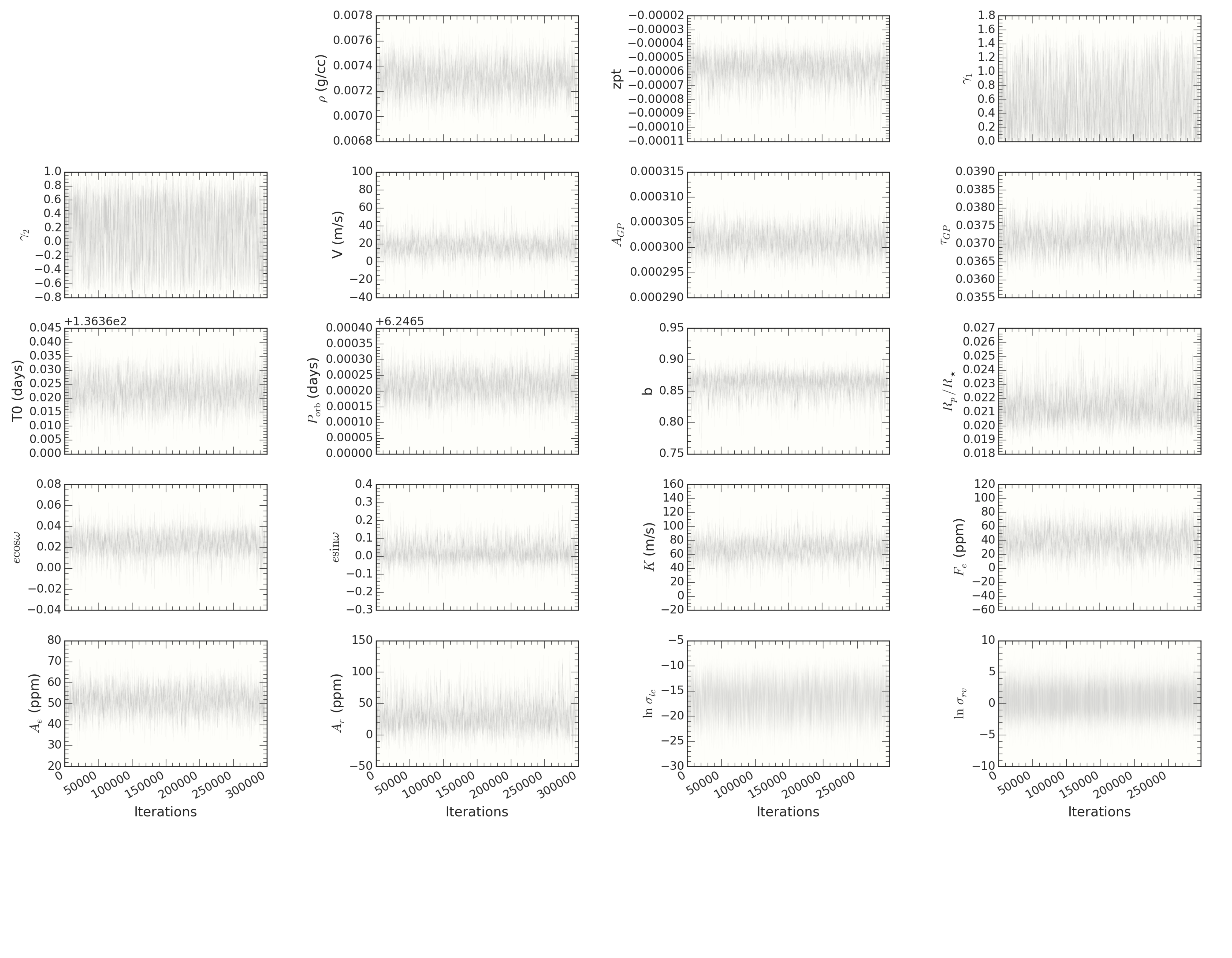}
\caption{The MCMC chains of the 19 individual parameters included in our model. We show here every 10th sample in order to make the plot easier to parse.}
\label{fig:chains}
\end{figure}

\begin{figure}
\includegraphics[width=0.97\textwidth]{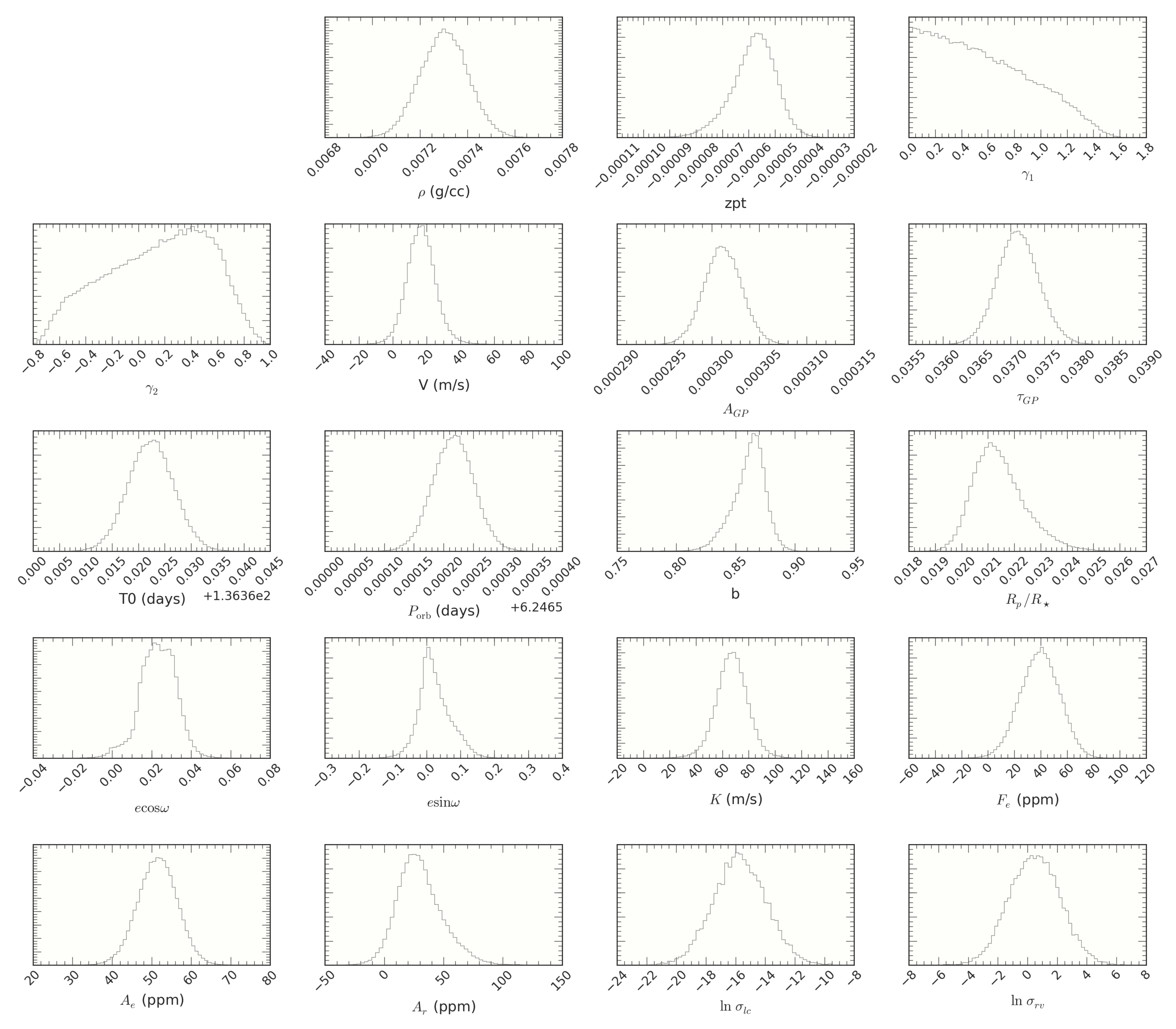}
\caption{Histograms of the marginalized posteriors of the 19 parameters included in our MCMC planet model.}
\label{fig:chain_hist}
\end{figure}

In this section we have included a number of additional plots to allow the reader to better assess the convergence of our MCMC chains. In Figure~\ref{fig:chains} we show a trace of each of the 19 parameters included in our model, while Figure~\ref{fig:chain_hist} shows histograms representing marginalized posteriors of the modeled parameters.

\end{appendix}
\end{document}